\input{aipcheck}

\newcommand{\eqa} {\begin{eqnarray} }
\newcommand{\eqe} {\end{eqnarray}}
\newcommand{\beq} {\begin{equation}}
\newcommand{\eeq} {\end{equation}}

\documentclass[
    ,final            % use final for the camera ready runs
  ]
  {aipproc}

\layoutstyle{6x9}

\begin{document}

\title{Rapidity Gap Events for Squark Pair Production at the LHC}

\classification{11.30.Pb, 12.60.Jv, 14.80.Ly \vspace{-4.0mm}}
\keywords      {Supersymmetry, Rapidity Gaps, LHC}

\author{Sascha Bornhauser}{
  address={ PandA, University of New
  Mexico, 800 Yale Blvd NE Albuquerque, NM 87131, USA}
}

\author{Manuel Drees}{
  address={Physikalisches Institut and BCTP, Universit\"at Bonn, Nussallee 12, D53115 Bonn, Germany}
  ,altaddress={KIAS, School of Physics, Seoul 130--012, Korea}
}

\author{Herbert K. Dreiner}{
  address={Physikalisches Institut and BCTP, Universit\"at Bonn, Nussallee 12, D53115 Bonn, Germany}
}

\author{Jong Soo Kim}{
  address={Institut f\"ur Physik, Technische Universit\"at Dortmund,
  D-44221 Dortmund, Germany} % additional visiting address
}

\begin{abstract}
  The exchange of electroweak gauginos in the $t-$ or $u-$channel allows
  squark pair production at hadron colliders without color exchange between
  the squarks. This can give rise to events where little or no energy is
  deposited in the detector between the squark decay products.  We discuss the
  potential for detection of such rapidity gap events at the Large Hadron
  Collider (LHC). 
  We present an analysis with full event simulation using
  PYTHIA as well as Herwig++, but without detector simulation. We analyze the
  transverse energy deposited between the jets from squark decay, as well as
  the probability of finding a third jet in between the two hardest jets. For
  the mSUGRA benchmark point SPS1a we find statistically significant
  evidence for a color singlet exchange contribution. 
\end{abstract}

\maketitle

%%%%%%%%%%%%%%%%%%%%%%%%%%%%%%%%%%%%%%%%%%%%
%% MAINMATTER
%%%%%%%%%%%%%%%%%%%%%%%%%%%%%%%%%%%%%%%%%%%%
\vspace{-4.0mm}
\section{SUSY Rapidity Gaps}
\vspace{-4.0mm}
One of the main objectives of the Large Hadron Collider (LHC) is the
search for supersymmetric (SUSY) particles \cite{susyrev}. In the
energy range of the LHC we expect squark pair production to be one of
the most important channels for the production of superparticles
\cite{qcdlo}. 
Squark pair production includes contributions with electroweak (EW)
exchange particles at tree \cite{oldew,bddk1} or one-loop \cite{nlo_ew} level.
The tree--level EW contributions can change the production cross section by up
to $50\%$ \cite{bddk1}. Moreover, EW gaugino exchange in the $t-$ or
$u-$channel gives rise to events with no color connection between the produced
squarks. QCD radiation then preferentially takes place in the phase space
region between the respective color connected initial quark and final squark,
not between the two outgoing squarks. If the rapidity region between both
squarks is indeed free of QCD radiation it is called a ``rapidity gap''. The
situation is different for the lowest order QCD contribution. The final
squarks are color--connected and radiation into the region between them is
expected. This difference might allow to isolate events with electroweak
gaugino (color singlet [CS]) exchange, which could e.g. lead to new methods to
determine their masses and couplings.
The above discussion describes a single partonic reaction producing stable
squarks. In reality, the squarks will decay. Even if we assume that each
squark decays into a single jet 
, the rapidity distribution of these jets will differ
from that of the squarks. Squark decay also leads to additional parton showers
from final state radiation. Moreover, the underlying event produced by the
beam remnants and their interactions can also deposit energy in the gap.
\vspace{-4.0mm}
\section{Numerical Simulation}
\vspace{-4.0mm}
In the following we perform a full event simulation, including squark decays, hadronization,
jet reconstruction and the ``underlying event'', however without simulating
the detector. To be specific, we consider rapidity--gap events for squark pair
production at the LHC, where electroweak (EW) contributions at tree level are
included. The production of the first two generations of squarks via $t-$ and
$u-$channel diagrams is taken into account. $s-$channel contributions are
neglected, since they are quite small \cite{bddk1}, and are not expected to
lead to rapidity--gap events. 
The mass
spectrum and branching ratios of the sparticles are obtained from SPheno
\cite{spheno}.  Analytical expressions for the squared and averaged matrix
elements are given in Ref.~\cite{bddk1}. We implemented the relevant matrix
elements for QCD and as well for EW contributions in a simple parton--level
simulation. Jets are reconstructed via the $k_T$ clustering algorithm of
FastJet \cite{Cacciari:2006sm}. Events are analyzed using the program
package root \cite{Brun:1997pa}.

We impose the following cuts: the two 
highest transverse momentum jets have to satisfy:
$ \label{etj}
E_T(j_i) \ge 100\,{\rm GeV} \, ; \quad \left|\eta(j_i)\right|
\leq 5.0 \quad (i=1,2).
$
We further suppress SM backgrounds by requiring a large amount of
missing transverse energy:
$ \label{etmiss}
E_T\hspace*{-4.5mm}/ \hspace*{4.5mm} \geq 100 \, {\rm GeV}.
$
Squark pair events containing at least one $SU(2)$ singlet squark, which give rise to quite small EW contributions \cite{bddk1}, are
suppressed by requiring the existence of two like--sign charged
leptons, with
$ \label{ptl}
p_T(\ell_i) \geq 5 \, {\rm GeV}\,; \quad \left| \eta(\ell_i) \right| \leq
2.4 \quad (i=1,2).
$
In order to be able to define a meaningful rapidity gap, the two
leading jets should be well separated in rapidity:
$ \label{etaj}
\Delta\eta \equiv \left| \eta(j_1) - \eta(j_2) \right| \ge3.0 \, .
$
We have to take into account that the two jets have finite radii. The
``gap region'' is therefore defined as: 
$\label{gap_region}
\rm{min}[\eta(j_1),\eta(j_2)] + 0.7 \le \eta \le
\rm{max}[\eta(j_1),\eta(j_2)] -0.7 \, .
$
One can expect that most of the particles produced during
hadronization are within the cone of $0.7$ of the corresponding jets \cite{BjorkenWW}. 
The overall efficiency of our cuts relative to the entire squark pair sample is
less than 1\%, cf.~\cite{Bornhauser:2009ru}.
Since we want to avoid ``event pile--up'', i.e. multiple $pp$ interactions
during the same bunch crossing, we assume an integrated luminosity of 40
fb$^{-1}$ at $\sqrt{s} = 14$ TeV. 
\begin{figure}
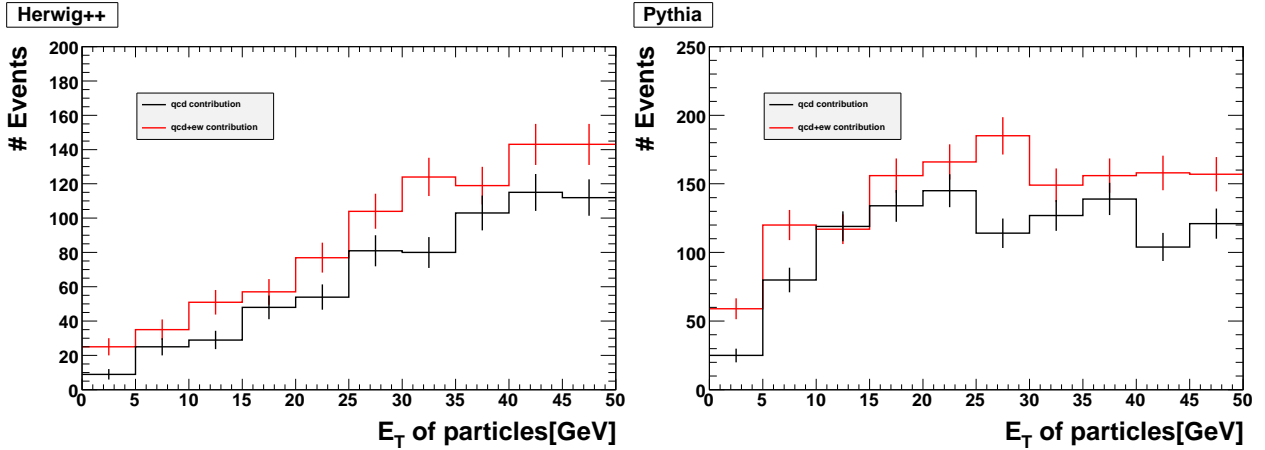

%\begin{center}
%\vspace{+0.7cm}
\rotatebox{270}{\includegraphics[scale=0.35]{plot4h.epsi}}
\rotatebox{270}{\includegraphics[scale=0.35]{plot4p.epsi}}
\caption{Transverse energy in the rapidity--gap region
  as predicted by full event simulations of squark
  pair production using HERWIG++ and PYTHIA 6.4.  Black
  histograms show pure QCD results, while the red (gray) histograms
  include electroweak contributions. The errors are statistical only.\vspace{-5.0mm}}
\label{fig1}
%\end{center}
\end{figure}

\begin{figure}[h!]
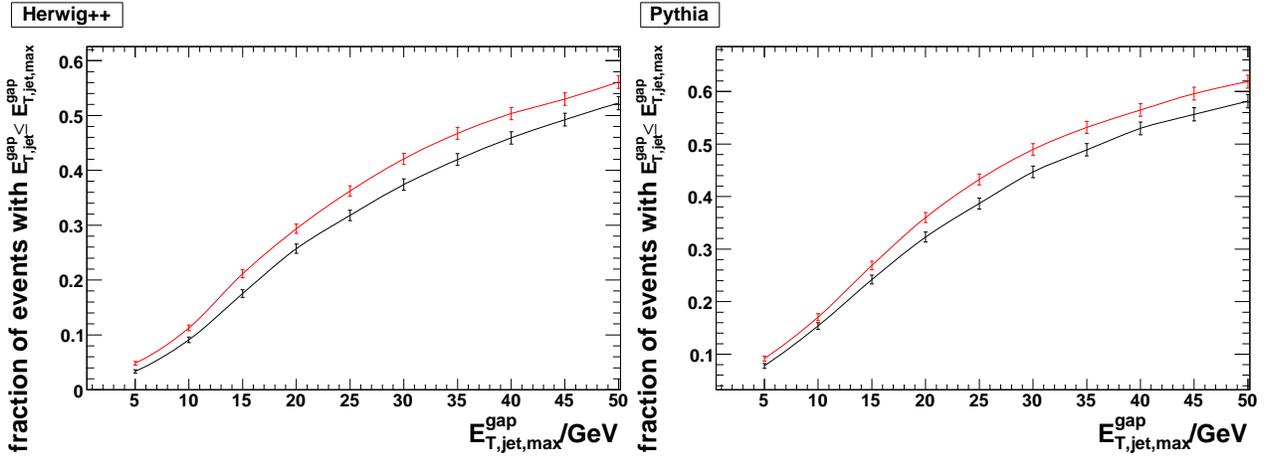
 
%\begin{center}
\rotatebox{270}{\includegraphics[scale=0.35]{plot3h.epsi}}\\
\vspace*{8mm}
\rotatebox{270}{\includegraphics[scale=0.35]{plot3p.epsi}}
\caption{Fraction of squark pair events passing a minijet veto in the
  rapidity--gap region, as predicted using full
  event simulations using Herwig++ and PYTHIA~6.4.  The
  black curve is for the pure QCD sample, and the red (gray) curve for
  the QCD+EW sample.\vspace{-5.0mm}}
\label{fig4}
%\end{center}
\end{figure}
\medskip

Our first attempt to isolate ``rapidity gap events'' uses a completely
inclusive quantity. We define $E_{T,{\rm particles}}^{\rm{gap}}$ as the total
transverse energy deposited in the gap region; this is computed from all photons and hadrons in the
event (after hadronization and decay of unstable hadrons), but does {\em not}
include the leptons produced in $\tilde \chi^0$ and $\tilde\chi^\pm$ decays.
The distribution of $E_{T,{\rm particles}}^{\rm{gap}}$ is shown in
Fig.~\ref{fig1} for Herwig++ and PYTHIA~6.4. In this and all
following figures, black and red histograms denote pure QCD and QCD+EW
predictions, respectively. We also show the statistical error for each bin.

We note that including EW contributions increases the number of
events, although in most bins this effect is statistically not very
significant. However, in the first bin, where the total $E_T$ is less
than 5 GeV, the inclusion of these CS exchange contributions increases
the number of events by a factor of $2.8 \pm 1.1$ and $2.36 \pm 0.56$
in the Herwig++ and PYTHIA 6.4 simulations, respectively. This
indicates that CS exchange does lead to ``gap'' events where little or
no energy is deposited between the two hard jets.

The
difference between the two generators is as large as the effect from
the CS events: PYTHIA 6.4 without CS exchange contributions predicts
almost exactly the same number of events in the first bin as Herwig++
with CS exchange. PYTHIA 6.4 also predicts a $E_{T,{\rm particles}}^{\rm{gap}}$ 
distribution which is quite flat beyond 20
GeV, whereas the distribution predicted by Herwig++ flattens out only
at about 40 GeV. One might thus be able to use the higher bins, where
the effect of the CS exchange contributions is not very sizable, to
decide which generator describes the data better, or to tune the Monte
Carlo generators to the data. This should reduce the difference
between the two predictions.
\medskip

Predicting the total transverse energy flow is difficult, since this
observable is strongly affected by semi-- and non--perturbative
effects. 
Thus, we discuss in the following
the occurrence of relatively soft ``minijets'' in the gap region.
Fig.~\ref{fig4} shows the fraction of events where the energy
$E_{T,\rm{jet}}^{\rm{gap}}$ of the most energetic jet in the gap region is less than the value $E_{T,\rm{jet,max}}^{\rm{gap}}$
displayed on the $x-$axis. Since Fig.~\ref{fig4} shows event {\em fractions},
all curves asymptotically approach 1 at large $E_{T,\rm{jet,max}}^{\rm{gap}}$.
We assume that jets with transverse energy above $E_{T,{\rm thresh}} = 5$ GeV
can be reconstructed. If the true threshold is higher, the curves should
simply be replaced by constants for $E_{T,\rm{jet,max}}^{\rm{gap}} \leq
E_{T,{\rm thresh}}$. We expect that the underlying event by itself generates
few, if any, reconstructable jets. The results here are nevertheless still not
quite immune to non--perturbative effects, since reconstructed jets may also
contain a few particles stemming from the underlying event. Note also that a
jet whose axis lies in the gap region might contain (mostly quite soft)
particles that lie outside of this region. Conversely, even though we use a
cluster algorithm, where by definition each particle belongs to some jet, some
particles in the gap region might be assigned to a jet whose axis lies outside
this region.

We see that PYTHIA~6.4 predicts more events without jet
in the gap region than Herwig++. 
This is consistent with the observation that PYTHIA predicted more events with little
or no energy deposited in the gap region and tends to generate
fewer hard gluons than Herwig++ does, see \cite{Bornhauser:2009ru}. 
Moreover, both PYTHIA~6.4 and Herwig++ predict a significant increase
of the fraction of events without jet in the gap region once EW, CS
exchange contributions are included; the effect is statistically most
significant for $E_{T,\rm{jet,max}}^{\rm{gap}} \sim 20$ to 40
GeV. Here both generators predict an increase of the fraction of
events without (sufficiently hard) jet in the gap by about
0.05. Taking a threshold energy of 30 GeV as an example, Herwig++
predicts about 1,570 out of the total of 4,032 events without jet in
the gap once EW effects are included. This should allow to measure the
fraction of events without jet in the gap to an accuracy of about
0.01; a shift of the jet--in--the gap fraction by 0.05 thus
corresponds to a change by about five standard deviations (statistical
error only). 

Unfortunately a pure SUSY QCD PYTHIA simulation leads to a fraction of
events without jet of 0.45 for the same threshold energy, which is
{\em higher} than the Herwig++ prediction {\em including} EW
contributions. Clearly these large discrepancies between the two MC
generators have to be resolved before reliable conclusions about the
color flow in squark pair events can be drawn.
In Ref.~\cite{Bornhauser:2009ru} we showed that a similar difference between the
predictions of the two generators also exists for the
jet--in--the--gap cross section in standard QCD. This indicates that
such standard QCD events can be used during the earliest phases of LHC
running to improve the event generators, hopefully to the extent that
the difference between their predictions becomes significantly smaller
than the effect from the EW contributions. 
\vspace{-6.0mm}
 \bibliographystyle{aipproc}   % if natbib is available
% %\bibliographystyle{aipprocl} % if natbib is missing
% 
% %%%%%%%%%%%%%%%%%%%%%%%%%%%%%%%%%%%%%%%%%%%
% %% You probably want to use your own bibtex database here
% %%%%%%%%%%%%%%%%%%%%%%%%%%%%%%%%%%%%%%%%%%%
% \bibliography{sample}
% 
% %%%%%%%%%%%%%%%%%%%%%%%%%%%%%%%%%%%%%%%%%%%
% %% Just a reminder that you may have to run bibtex
% %% All of it up to \end{document} can be removed
% %% if you don't like the warning.
% %%%%%%%%%%%%%%%%%%%%%%%%%%%%%%%%%%%%%%%%%%%
% \IfFileExists{\jobname.bbl}{}
%  {\typeout{}
%   \typeout{******************************************}
%   \typeout{** Please run "bibtex \jobname" to optain}
%   \typeout{** the bibliography and then re-run LaTeX}
%   \typeout{** twice to fix the references!}
%   \typeout{******************************************}
%   \typeout{}
%  }

\end{document}